\documentclass[review]{elsarticle}

\usepackage{lineno,hyperref,amsmath}
\modulolinenumbers[5]

\journal{Astroparticle Physics}

\usepackage{amssymb}
\begin{document}

\begin{frontmatter}

\title{Significance in Gamma Ray Astronomy with Systematic Errors}
%%\tnotetext[mytitlenote]{Fully documented templates are available in the elsarticle package on \href{http://www.ctan.org/tex-archive/macros/latex/contrib/elsarticle}{CTAN}.}
%%\tnotetext{Significance in Gamma-Ray Astronomy with a Systematic Error on the Exposure Ratio}

\author{Gerrit Spengler\footnote{gerrit.spengler@fysik.su.se}}
\address{Oscar Klein Centre for Cosmoparticle Physics, Physics Department, Stockholm University, Albanova University Centre, SE-10691 Stockholm}
%%\fntext[myfootnote]{Since 1880.}




\begin{abstract}
The influence of systematic errors on the calculation of the statistical significance of a $\gamma$-ray signal with the frequently invoked 
Li and Ma method is investigated. A simple criterion is derived to decide whether the Li and Ma method can be applied in the presence of 
systematic errors. An alternative method is discussed 
for cases where systematic errors are too large for the application of the original Li and Ma method. This alternative method reduces to the Li and Ma method when systematic errors are negligible. Finally, it is shown that the consideration of systematic errors will be important in many analyses of data from the planned Cherenkov Telescope Array.
\end{abstract}


\end{frontmatter}

%\linenumbers

\section{Introduction}
The calculation of the statistical significance of a measured signal is a central part of a data analysis in Very High Energy (VHE) $\gamma$-ray astronomy. In a typical analysis, the number of events, $N_\mathrm{ON}$, that are detected in a signal region is compared 
to the number of events, $\alpha N_\mathrm{OFF}$, that are expected in the signal region. The expected number of events in the signal region results from a measurement of $N_\mathrm{OFF}$ events in a dedicated background region. Different 
methods (see \cite{berge}) exist to construct the background region for a given signal region such that the event acceptances in the signal and the background region are equal. If the event acceptances in the signal and the background region are equal, 
the normalization factor $\alpha$ is equal to the ratio of the exposures in the signal and the background region.\\
The statistical significance of a $\gamma$-ray signal, $\Delta = N_\mathrm{ON} - \alpha N_\mathrm{OFF}$, is derived in \cite{lima} to be 
\begin{equation}
S_\mathrm{LiMa}(N_\mathrm{ON}, N_\mathrm{OFF}, \alpha)=\mathrm{sign}(\Delta)\sqrt{-2\left(N_\mathrm{ON}\ln(\frac{\overline{N_\mathrm{ON}}}{N_\mathrm{ON}})+N_\mathrm{OFF}\ln(\frac{\overline{N_\mathrm{OFF}}}{N_\mathrm{OFF}})\right)}
\label{lima_sig}
\end{equation}
where $\overline{N_\mathrm{ON}} = \alpha \overline{N_\mathrm{OFF}}$
and
\begin{equation}
\overline{N_\mathrm{OFF}}=\overline{N_\mathrm{OFF}}(\alpha)=\frac{N_\mathrm{ON}+N_\mathrm{OFF}}{1+\alpha}\:\mathrm{.}
\label{noff_lima}
\end{equation}
The normalization factor $\alpha$ is taken to have a negligible error, $\sigma_\alpha$, in the derivation of Eq. \ref{lima_sig} in \cite{lima}.\\
The order of magnitude of the statistical error on the $\gamma$-ray signal due to Poisson fluctuations of $N_\mathrm{ON}$ and $N_\mathrm{OFF}$ can be estimated to be $\Delta_\mathrm{Background}\approx\sqrt{N_\mathrm{ON}+\alpha^2 N_\mathrm{OFF}}$. When no signal is present, it holds $N_\mathrm{ON}\approx\alpha N_\mathrm{OFF}$ which leads to $\Delta_\mathrm{Background}\approx\sqrt{\alpha(\alpha+1)\:N_\mathrm{OFF}}$. 
A similar back of the envelope estimation for the error on the $\gamma$-ray signal, propagated from an error on the normalization factor, leads to 
$\Delta_\mathrm{\alpha} \approx\sigma_\alpha N_\mathrm{OFF}$.\\
The application of $S_\mathrm{LiMa}$ to calculate the statistical significance of a $\gamma$-ray event signal is justified if $\Delta_\mathrm{\alpha} \ll \Delta_\mathrm{Background}$. Using the back of the envelope estimation for the case where no signal events are measured in the signal region, the condition $\Delta_\mathrm{\alpha}\ll\Delta_\mathrm{Background}$ translates into the condition 
\begin{equation}
\frac{\sigma_\alpha}{\alpha} \ll\sqrt{\frac{1+\alpha}{\alpha}\:\frac{1}{ N_\mathrm{OFF}}}
\label{envelope}
\end{equation}
for the relative error on the normalization factor.\\
The following discussion focuses on data acquired with imaging atmospheric Cherenkov telescopes (IACTs). However, $S_\mathrm{LiMa}$ is also frequently applied in analyses of data from ground based water Cherenkov telescopes (see e.g. \cite{milagro} and 
\cite{milagro2}) and extensive air shower arrays (see e.g. \cite{auger}).\\
The High Energy Stereoscopic System (H.E.S.S.) is an array of IACTs that operates in the Namibian Khomas Highland since $2003$. 
Observations of the Crab nebula with H.E.S.S. result in the detection of approximately $100$ background events that pass standard Hillas $\gamma$-ray event selection criteria 
per $30$ min observation time for a normalization factor of $\alpha = 0.2$ (see \cite{crab} for details). 
Similar background event rates hold for analyses of data from observations of point like $\gamma$-ray sources with other current generation IACT arrays such as MAGIC \cite{magicCrab} and VERITAS \cite{veritas}. According to Eq. \ref{envelope}, the relative error on the normalization factor must be much smaller than $\sigma_\alpha/\alpha \approx 25\%$ when $S_\mathrm{LiMa}$ is used to calculate the significance of a $\gamma$-ray signal with $N_\mathrm{OFF}\approx 100$ and $\alpha=0.2$. More precisely, it will be shown later that in this case $\sigma_\alpha/\alpha$ must be known to about $3\%$ to justify the application of $S_\mathrm{LiMa}$.\\
For the planned Cherenkov Telescope Array (CTA, see \cite{CTA}), the increase in the number of telescopes will (compared to current generation IACTs) lead to a factor of $10$ larger effective area. An increase of the background event rate for CTA of a similar factor of $10$ (again compared to current generation IACT arrays) is expected from the enlarged effective area (see \cite{ctaMC}). Consequently, approximately $10^3$ background events per $30$ min observation time are expected in a typical point source analysis of CTA data with $\alpha=0.2$. In this case, it is estimated with Eq. \ref{envelope} that the relative error on the normalization factor must be much smaller than $8\%$. Otherwise, an error on the normalization factor must be considered. Again more precisely, it will be shown later that the relative error on the normalization factor must be known to about one order of magnitude better than $8\%$ for the application of $S_\mathrm{LiMa}$ in this situation.\\
In analyses of extended $\gamma$-ray sources (e.g. Supernova Remnants, \cite{velaJr}) or galactic dark matter searches (e.g. \cite{daniil}), the increased size of the signal region leads easily to background event rates which are an order of magnitude larger than for point source analyses. It is obvious that in those cases, the normalization factor $\alpha$ must be known with an even better precision than in point source analyses.\\
It is arguable whether the normalization factor is in general known with the precision that is required for the application of $S_\mathrm{LiMa}$. This holds in particular for analysis of data from the planned CTA experiment.\\
This paper extends the method for the calculation of the statistical significance of a $\gamma$-ray signal first proposed in \cite{lima} to include an error on the normalization factor. In addition to other authors, who discussed the same problem (e.g. \cite{hugh}, \cite{klepser}), a simple expression for the calculation of the significance of a measured $\gamma$-ray signal is derived. Moreover, the criterion given by Eq. \ref{envelope} to decide whether a given error on the normalization factor must be considered in the calculation of the statistical significance of a $\gamma$-ray signal is tested in Monte Carlo simulations.\\
The structure of the paper is as follows: The effect of the neglection of a systematic error on the normalization factor on $S_\mathrm{LiMa}$ is 
quantitatively discussed in section \ref{limaError}. A modified expression for the significance calculation, which considers an error on the normalization factor, is derived in section \ref{modified}.\\
Section \ref{combin} discusses more specifically the influence of systematic errors on the calculation of statistical significances in VHE $\gamma$-ray astronomy with IACTs like H.E.S.S. or CTA.

\section{The Li and Ma Significance with a Gaussian Distributed Normalization Factor}
\label{limaError}
\begin{figure}{}
\centering
\includegraphics*[scale=0.62]{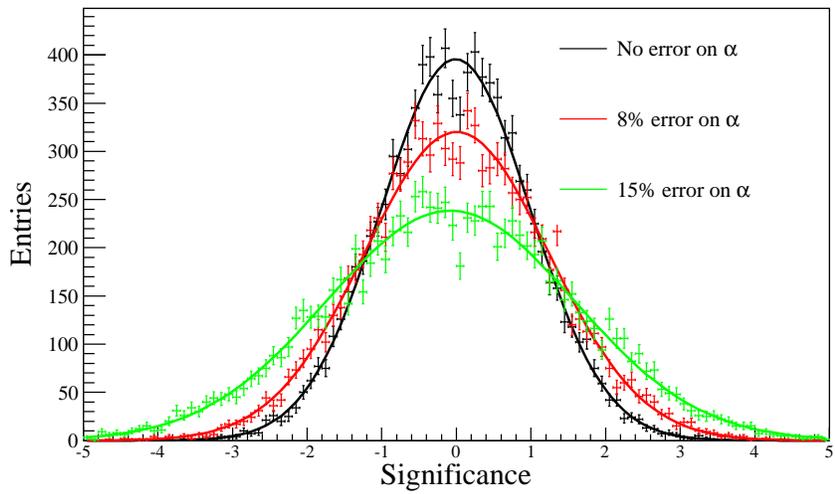}
\caption{Distribution of $S_\mathrm{LiMa}$ (Eq. \ref{lima_sig}) when $N_\mathrm{ON}=\mathrm{Pois}(\alpha b)$ and $N_\mathrm{OFF}=\mathrm{Pois}(b)$ are independently Poisson distributed with $b=500$ events. 
The normalization factor, $\alpha$, is distributed like a Gaussian with mean $0.2$ and relative width according to the legend. The width of the Gaussian fit to the significance distribution is compatible with being one if the error on the normalization factor is 
vanishing. However, the width of the Gaussian fit to the significance distribution increases with the relative error on the normalization factor.}
\label{lima_distribution}
\end{figure}

\begin{figure}{}
\centering
\includegraphics*[scale=0.62]{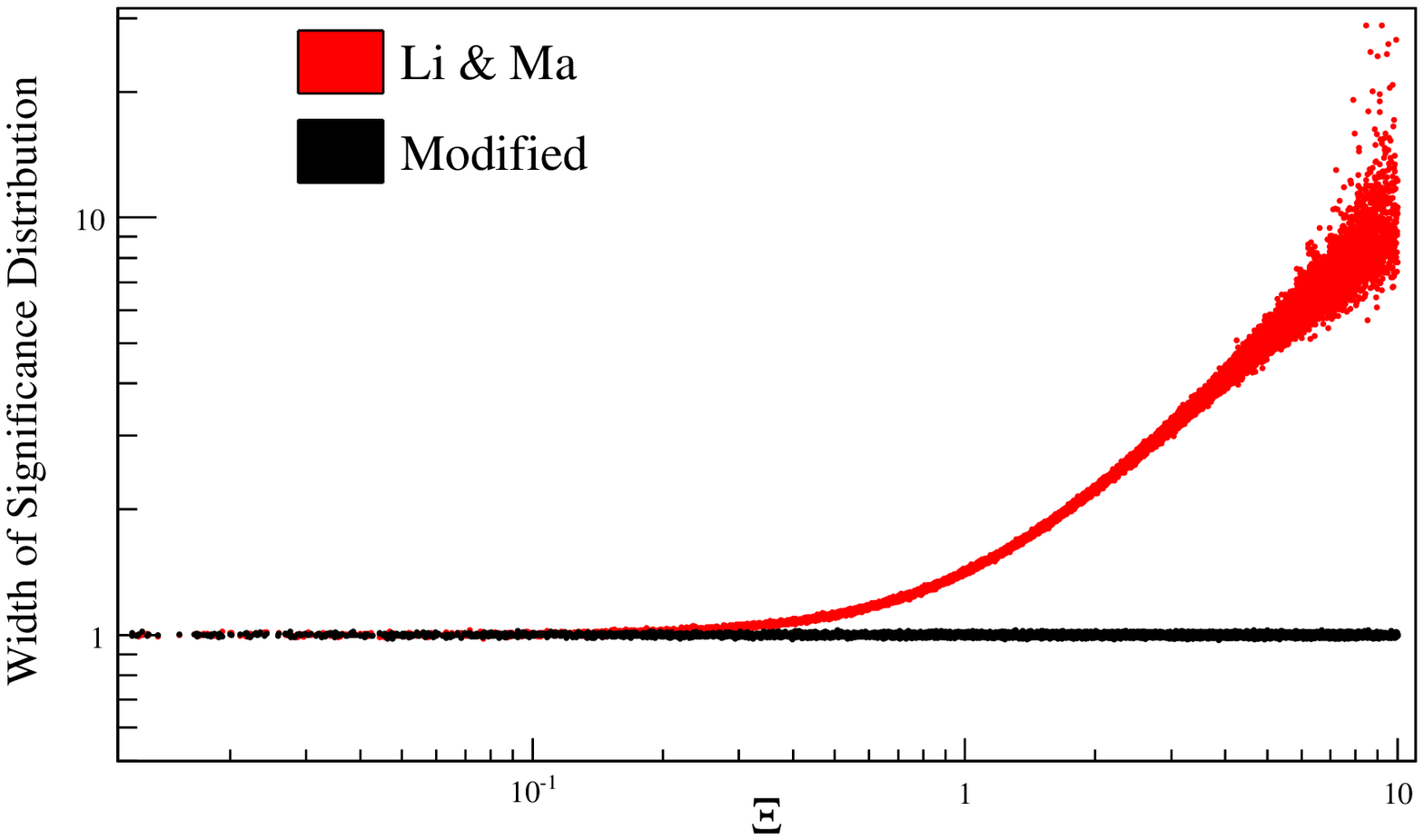}
\caption{Fitted Gaussian width of the distributions of the significance as calculated with $S_\mathrm{LiMa}$ (Eq. \ref{lima_sig}, 
red points) and $S_\mathrm{Modified}$ (Eq. \ref{modified_sig}, black points) as a function of $\Xi$ (see Eq. \ref{xi}). The parameters $N_\mathrm{OFF}$, $\alpha$ and $\sigma_\alpha/\alpha$ are randomly distributed within 
$N_\mathrm{OFF}\in [100, 10000]$, $\alpha\in [0.1,2]$ and $\sigma_\alpha/\alpha\in [0,15\%]$.}
\label{lima_width}
\end{figure}
Figure \ref{lima_distribution} shows the distribution of the significance as calculated with $S_\mathrm{LiMa}$ 
when $N_\mathrm{ON}=\mathrm{Pois}(\alpha b)$ and $N_\mathrm{OFF}=\mathrm{Pois}(b)$ are independently Poisson distributed with $b=500$ events. No signal events are simulated and it is expected that the Gaussian fit of the distribution of the significances results in a distribution which is compatible with being a standard normal. The distribution of the significance is shown in Fig. \ref{lima_distribution} for a fixed normalization factor, $\alpha = 0.2$. Additionally shown is the distribution of $S_\mathrm{LiMa}$ for two Gaussian distributed normalization factors. The mean of the assumed distributions of the normalization factor is $\alpha = 0.2$ in both cases. Respectively, the relative widths of the Gaussian distributions for the normalization factor are $\sigma_\alpha/\alpha = 8\%$ and $\sigma_\alpha/\alpha = 15\%$. The assumption of a Gaussian distributed normalization factor is here and in the following restricted to small relative errors on the normalization factor ($\sigma_\alpha/\alpha\lesssim 15\%$). For large relative errors on the normalization factor, the distribution cannot in general be assumed to be Gaussian, e.g. because the normalization factor must be larger than zero for physical reasons. However, the assumption of small relative errors on the normalization factor is reasonable because large relative errors can easily be identified and corrected for in analyses.\\
The mean of all fitted significance distributions that are shown in Fig. \ref{lima_distribution} is compatible with zero. 
However, the width of the Gaussian fit to the significance distribution is only compatible with being one when the error on the normalization factor vanishes. For non-zero relative errors on the normalization factor, the width of the Gaussian fit to the significance distribution increases with the relative error on the normalization factor. 
In other words, the absolute value of the significance is overestimated by $S_\mathrm{LiMa}$ 
when the relative error on the normalization factor is not vanishing.\\
The examples in Fig. \ref{lima_distribution} hold only for special values for the mean number of background events and the normalization factor. In general, the criterion for the applicability of $S_\mathrm{LiMa}$ given in Eq. \ref{envelope} translates into the condition
\begin{equation}
\Xi = \sigma_\alpha\: \sqrt{\frac{N_\mathrm{OFF}}{\alpha(\alpha+1)}} \ll 1\:\mathrm{.}
\label{xi}
\end{equation}
Figure \ref{lima_width} shows the width of the Gaussian fit to the distribution of the significance as a function of $\Xi$. Here, the parameter $\Xi$ is calculated for a random selection of 
$\alpha$, $\sigma_\alpha$ and $N_\mathrm{OFF}$. The simulated number of background events, $N_\mathrm{OFF}$, is 
Poisson distributed with uniformly distributed means in between $100$ and $10000$. The normalization factor, $\alpha$, is Gaussian distributed with uniformly distributed means in between $0.1$ and $2$ and with uniformly distributed relative widths 
in between $\sigma_\alpha/\alpha=0$ and $\sigma_\alpha/\alpha=15\%$. Figure \ref{lima_width} shows that the width of the Gaussian fit to the distribution of $S_\mathrm{LiMa}$ is indeed larger than one if Eq. \ref{xi} does not hold. More specifically, Fig. \ref{lima_width} shows that the width of the Gaussian fit to the distribution of $S_\mathrm{LiMa}$ becomes larger than one if $\Xi\gtrsim 0.1$.\\
$S_\mathrm{LiMa}$ is, under the assumption of a vanishing error on the normalization factor, constructed in \cite{lima} to be standard normal distributed when the null hypothesis is true, i.e. when no signal events are present. In turn, $S_\mathrm{LiMa}$ is not applicable when the systematic error on the normalization factor causes a distribution of $S_\mathrm{LiMa}$ which is not compatible with being a standard normal distribution although the null hypothesis is true. This means that $S_\mathrm{LiMa}$ is applicable if 
\begin{equation}
\Xi\lesssim 0.1
\label{final_xi}
\end{equation}
is fulfilled.\\
Equation \ref{final_xi} translates into the condition $\sigma_\alpha/\alpha\lesssim 0.1\:\sqrt{(\alpha+1)/(\alpha\,N_\mathrm{OFF})}$ for the relative error on the normalization. This condition is more precise than  Eq. \ref{envelope}. Also, this condition was already used in the introduction when the maximal relative error on the normalization factor that allows the application of $S_\mathrm{LiMa}$ in typical point source analyses with current IACTs and the planned CTA detector was estimated.\\
For cases where a systematic error on the normalization factor cannot be neglected, a modified expression for the calculation of the statistical significance of 
a $\gamma$-ray signal is derived in the next section.

\section{Modified Significance for a Gaussian Distributed Normalization Factor}
\label{modified}
Consider, similar to the approach in \cite{lima}, the likelihood function 
\begin{equation}
\begin{aligned}
L&(N_\mathrm{ON}, N_\mathrm{OFF}, \alpha | s, b, \alpha^*)  \\
&= \mathrm{Pois}(N_\mathrm{ON},\alpha^* b + s)\,\mathrm{Pois}(N_\mathrm{OFF},b)\, \mathrm{Gaus}(\alpha^*,\alpha,\sigma_\alpha)\:\mathrm{.}
\end{aligned}
\label{likeli_gen}
\end{equation}
Again, this likelihood function is in this work assumed to describe the case of small relative errors on the normalization factor, i.e. $\sigma_\alpha/\alpha\lesssim 15\%$, to avoid e.g. negative $\alpha^*$. The general likelihood function given by Eq. \ref{likeli_gen} has already been investigated in the literature (see \cite{hugh}, \cite{conrad}, \cite{klepser}). However, no simple result for the statistical significance of a measured signal under the assumption of Eq. \ref{likeli_gen} was derived before. Also, no simple criterion to decide whether Eq. \ref{likeli_gen} should be applied was stated before. This is done in the following.\\
The likelihood-ratio test statistic is used with the likelihood function given by Eq. \ref{likeli_gen} to compare the null hypothesis, i.e. that no signal is measured ($s=0$), with the alternative hypothesis ($s\neq 0$). The likelihood for the null hypothesis is 
\begin{equation}
L_0 = \mathrm{Pois}(N_\mathrm{ON},\alpha^* b)\, \mathrm{Pois}(N_\mathrm{OFF},b)\, \mathrm{Gaus}(\alpha^*,\alpha, \sigma_\alpha)
\label{lik0}
\end{equation}
and the two nuisance parameters $b$ (i.e. the mean number of background events) and $\alpha^*$ (i.e. the Gaussian distributed normalization factor with mean $\alpha$) 
are eliminated with the profile likelihood method (see e.g. \cite{rolke}). The profile likelihood condition $\partial L_0/\partial b = 0$ leads to $b=\overline{N_\mathrm{OFF}}(\alpha^*)$ where 
$\overline{N_\mathrm{OFF}}(\alpha^*)$ is defined in Eq. \ref{noff_lima}. The second profile likelihood condition, $\partial L_0/\partial\alpha^* = 0$, leads to the cubic expression 
\begin{equation}
\Lambda(\alpha^*)=\alpha^{*3} - \alpha^{*2} (\alpha-1) - \alpha^*(\alpha - \sigma_\alpha^2 N_\mathrm{OFF}) - \sigma_\alpha^2 N_\mathrm{ON} = 0\:\mathrm{.}
\label{alpha_profile1}
\end{equation}
Equation \ref{alpha_profile1} can have up to three real solutions. Since, for physical reasons, $\sigma_\alpha^2 N_\mathrm{ON}\geq 0$, it follows that $\Lambda(0)\leq 0$. Together with $\lim_{\alpha^*\rightarrow\infty}\Lambda(\alpha^*)=\infty$, 
it follows that there is always at least one real and positive solution to Eq. \ref{alpha_profile1}. For example, if $\sigma_\alpha=0$, Eq. \ref{alpha_profile1} has three solutions at $\alpha^*=-1$, $\alpha^*=0$ and $\alpha^*=\alpha$. 
Since $L_0$ is not defined for $\alpha^*=-1$ and $L_0=0$ for $\alpha^*=0$, the solution $\alpha^*=\alpha$ maximizes the likelihood for the null hypothesis.\\
When $(\alpha+1)^2>3\sigma_\alpha^2 N_\mathrm{OFF}$, the cubic equation $\Lambda(\alpha^*)=0$ has 
in general a local maximum at $\alpha^-$ and a local minimum at $\alpha^+$ where
\begin{equation}
\alpha^{\pm}=\frac{1}{3}\left(\alpha -1 \pm\sqrt{(\alpha+1)^2-3\sigma_\alpha^2 N_\mathrm{OFF}}\right)\:\mathrm{.}
\end{equation} 
If additionally $\mathrm{sign}(\Lambda(\alpha^+)\Lambda(\alpha^-))\neq 1$, there are multiple real solutions to Eq. \ref{alpha_profile1}. In this case, a numerical method can be employed to search for respectively one solution to Eq. \ref{alpha_profile1} in the intervals $(0,\alpha^-]$, $(\alpha^-,\alpha^+]$ and $(\alpha^+, \infty)$. Out of the up to three found solutions, the one solution which maximizes the likelihood given by Eq. \ref{lik0} is to be selected. However, for numerical reasons, it is in practice better to chose the solution to Eq. \ref{alpha_profile1} which maximizes
\begin{equation}
\begin{aligned}
&\ln\left(N_\mathrm{ON}!\,N_\mathrm{OFF}!\,L_0+N_\mathrm{ON}+N_\mathrm{OFF}\right)\\
&=N_\mathrm{ON}\ln(\alpha^*)+\left( N_\mathrm{ON}+N_\mathrm{OFF}\right)\ln\left(\frac{N_\mathrm{ON}+N_\mathrm{OFF}}{\alpha^*+1}\right)-\frac{(\alpha^*-\alpha)^2}{2\sigma_\alpha^2}\:\mathrm{.}
\end{aligned}
\end{equation}
If Eq. \ref{alpha_profile1} does not have local extrema or $\mathrm{sign}(\Lambda(\alpha^+)\Lambda(\alpha^-))=+1$, there is only one real solution to Eq. \ref{alpha_profile1} which can be found numerically in the interval $(0,\infty)$.\\
The parameters $s = N_\mathrm{ON} - \alpha^* b$, $b = N_\mathrm{OFF}$ and $\alpha^*=\alpha$ maximize the likelihood function (Eq. \ref{likeli_gen}) and lead to the likelihood 
\begin{equation}
L_1 = \mathrm{Pois}(N_\mathrm{ON},N_\mathrm{ON})\, \mathrm{Pois}(N_\mathrm{OFF},N_\mathrm{OFF})\, \frac{1}{\sigma_\alpha \sqrt{2\pi}}
\end{equation}
for the alternative hypothesis. The alternative hypothesis has one more parameter ($s$) than the null hypothesis. 
Following Wilks' theorem \cite{wilksTheorem}, it is assumed that the likelihood ratio test statistic $\mathrm{TS} = -2 \ln (L_0/L_1)$ 
is approximately distributed like a $\chi^2$-distribution with one degree of freedom. The validity of this assumption is justified with Monte Carlo simulations. The evaluation of the likelihood ratio test statistic $\mathrm{TS} = -2 \ln (L_0/L_1)$ gives the modified significance
\begin{equation}
\begin{aligned}
\mathrm{S}_\mathrm{Modified}&(N_\mathrm{ON}, N_\mathrm{OFF}, \alpha, \sigma_\alpha) =  \mathrm{sign}(N_\mathrm{ON}-\alpha N_\mathrm{OFF}) \sqrt{ \mathrm{TS} }\\
& =\mathrm{sign}(N_\mathrm{ON}-\alpha N_\mathrm{OFF}) \sqrt{ S^2_\mathrm{LiMa}(N_\mathrm{ON},N_\mathrm{OFF},\alpha^*) + \left( \frac{\alpha^* - \alpha}{\sigma_\alpha}\right)^2} \:\mathrm{.}
\end{aligned}
\label{modified_sig}
\end{equation}
Note that $S_\mathrm{Modified}\rightarrow S_\mathrm{LiMa}$ when $\alpha^*\rightarrow\alpha$, i.e. when $\sigma_\alpha\rightarrow 0$. This means that $S_\mathrm{Modified}$ reduces to $S_\mathrm{LiMa}$ when systematic errors are negligible.\\
Equation \ref{modified_sig} is tested in a parameter range ($N_\mathrm{OFF}\in [100,10000]$, $\alpha\in [0.1,2]$ and $\sigma_\alpha/\alpha\in [0,15\%]$) which is in practice relevant for analyses of data from IACTs. The result is shown in Fig. \ref{lima_width}. The distribution of $S_\mathrm{Modified}$ 
is compatible with being standard normal if $N_\mathrm{ON}=\mathrm{Pois}(\alpha b)$ and $N_\mathrm{OFF}=\mathrm{Pois}(b)$ are independently Poisson distributed with fixed $b$ for all tested combinations of $\alpha$, $N_\mathrm{OFF}$ and $\sigma_\alpha$. 
In particular, the width of the Gaussian fit to the distribution of $S_\mathrm{Modified}$ shown in Fig. \ref{lima_width} as a function of $\Xi$ is compatible with being one.\\
It is concluded that $S_\mathrm{Modified}$ is standard normal distributed when the null hypothesis is true and systematic errors are too large for the application $S_\mathrm{LiMa}$, i.e. if $\Xi\gtrsim 0.1$. For negligible systematic errors, $S_\mathrm{Modified}$ reduces to $S_\mathrm{LiMa}$. The range of possible application for $S_\mathrm{Modified}$ is thus larger than for $S_\mathrm{LiMa}$.

\section{Example: Influence of Systematic Errors in IACT Analyses}
\label{combin}
\begin{figure}{}
\centering
\includegraphics*[scale=0.62]{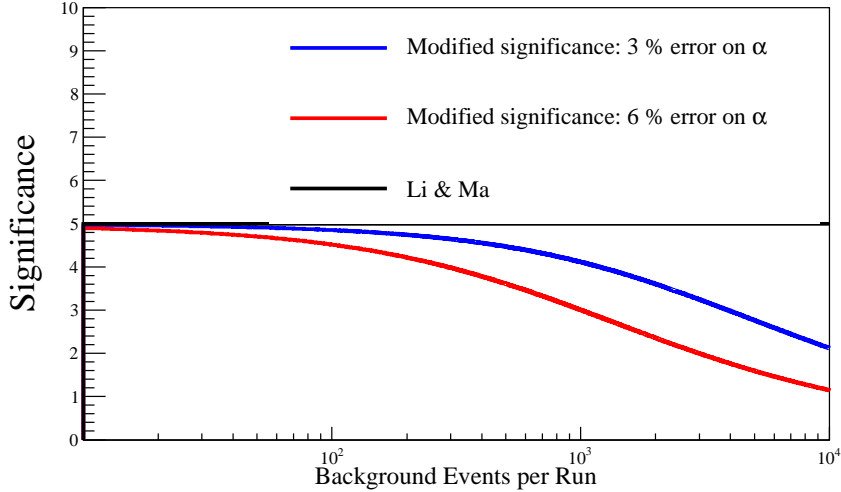}
\caption{Comparison of $S_\mathrm{LiMa}$ and $S_\mathrm{Modified}$ as a function of the number of background events per observation run for $\sigma_\alpha/\alpha=3\%$ (blue), $\sigma_\alpha/\alpha=6\%$ (red) and $\alpha=1$. The number of signal events is chosen such that $S_\mathrm{LiMa}=5.$}
\label{influence}
\end{figure}

As an illustration for the relevance of the consideration of systematic errors in the calculation of signal significances, consider the case of an IACT. Frequently (see e.g. \cite{berge}, \cite{crab}, \cite{daniil}), a background region 
for a given signal region is constructed in IACT data analyses under the assumption that the background event acceptance is rotationally symmetric around the observation position. In this case, the normalization factor $\alpha$ is the ratio of the exposures of the 
signal and background region. However, imbalances of the event acceptances between the signal and background region lead in general to a deviation between the normalization factor and the exposure ratio. 
Typical reasons for the imbalances between the event acceptances in the signal and the background region can e.g. be atmospheric differences, bright stars, varying night sky brightness or electronic problems. 
An order of magnitude for the deviation of the event acceptance from rotational symmetry 
around the observation position in an observation with H.E.S.S. is estimated in \cite{berge} to be $3\%$. Figure \ref{influence} shows $S_\mathrm{Modified}$ for an error of $3\%$ on the normalization factor as a function of the 
number of background events per observation run when the number of signal events is chosen such that $S_\mathrm{LiMa} = 5$.
It is evident that in a typical H.E.S.S. point source analysis with approximately $100$ background events per observation run (see \cite{crab}), the influence of expected deviations from rotational symmetry on the calculation of the signal significance 
is almost negligible. However, it is also obvious that the influence of systematic errors becomes larger when the number of background events per observation run increases. This situation can easily occur in analyses of data from 
extended signal regions where a spatially larger background region is used than in point source analyses. 
As discussed in the introduction, approximately $10^3$ background events are expected in a typical point source analysis of data acquired with the planned CTA experiment (\cite{CTA}, \cite{ctaMC}). Figure \ref{influence} shows clearly that this increase of the expected number of background events per observation run can decrease the source detection potential of CTA when analyses are performed under conditions that are otherwise similar to e.g. H.E.S.S. data analyses.\\
Typically, multiple ($i=1\ldots K$) observations of a signal region are combined in an IACT analysis such that the exposure ratio $\alpha_i$ is the same for all observations, i.e. $\alpha_i=\alpha$. However, deviations between the exposure ratio and the normalization factor from observation run to observation run lead to a distribution of normalization factors with mean $\alpha$. Again, the distribution of normalization factors is assumed to be Gaussian with width $\sigma_\alpha$. For the combined dataset, the resulting error on the mean normalization factor is $\sigma_\alpha/\sqrt{K}$. The total number of signal ($N_\mathrm{ON} = \sum_i N_\mathrm{ON,i}$) and the total number of background events ($N_\mathrm{OFF} = \sum_i N_\mathrm{OFF,i}$) 
are used together with the exposure ratio $\alpha$ as normalization factor to calculate the significance of the total $\gamma$-ray event signal. Consider the mean number of background events, $\langle N_\mathrm{OFF}\rangle = 1/K\sum_i N_\mathrm{OFF,i}$. The criterion given by Eq. \ref{final_xi} for the application of $S_\mathrm{LiMa}$ to data from one observation becomes
\begin{equation}
\Xi_\mathrm{Combined} = \frac{\sigma_\alpha}{\sqrt{K}}\,\sqrt{\frac{N_\mathrm{OFF}}{\alpha(\alpha+1)}} = \sigma_\alpha\,\sqrt{\frac{\langle N_\mathrm{OFF}\rangle}{\alpha(\alpha+1)}} \lesssim 0.1
\label{xi_comb}
\end{equation}
when data from multiple observations are combined. Equation \ref{xi_comb} means that given a normalization factor and an error on the normalization factor, the mean number of background events per observation is relevant to decide whether $S_\mathrm{LiMa}$ is applicable or not when multiple observations are combined. If Eq. \ref{xi_comb} is not fulfilled, Eq. \ref{modified_sig} can be used to calculate the statistical significance of the combined signal as $S_\mathrm{Modified}(N_\mathrm{ON},\, N_\mathrm{OFF},\,\alpha,\,\sigma_\alpha/\sqrt{K})$.\\
The argumentation given above doesn't cover an important exception that concerns the case where an analysis is set up such that the exposure ratios are not equal for data from differing observations. This situation can 
in general only be treated with a likelihood function that is binned in the exposure ratios (see \cite{hugh}) or even binned in more parameters that characterize the observation conditions (see \cite{klepser}).\\
The considerations in this section show that it is, in particular for the planned CTA experiment, important to estimate the typical systematic error on the normalization factor for an observation run in an IACT data analysis. Depending on the mean number of background events per 
observation run, the exposure ratio and the estimated error on the exposure ratio, the usage of $S_\mathrm{Modified}$ is preferred to the usage of $S_\mathrm{LiMa}$ for the calculation of the statistical significance of a $\gamma$-ray signal.

\section{Conclusion}
The influence of systematic fluctuations of the normalization factor $\alpha$ on the calculation of the statistical significance of a $\gamma$-ray signal was investigated. The discussion focussed on small relative fluctuations ($\sigma_\alpha/\alpha\lesssim 15\%$) of $\alpha$ around its expected value. It was shown that the Li and Ma method derived in \cite{lima} for the calculation of the statistical significance should only be applied to data acquired in one observation if Eq. \ref{final_xi} is fulfilled. Similarly, if data from multiple equations are combined, Eq. \ref{xi_comb} must be fulfilled or $S_\mathrm{LiMa}$ should not be applied. The absolute value of the statistical significance of a $\gamma$-ray signal is in general overestimated if Eq. \ref{final_xi} or, respectively, Eq. \ref{xi_comb} is not fulfilled. An alternative method (Eq. \ref{modified_sig}) for the calculation of the statistical significance was derived. This method takes random fluctuations of the normalization factor around its expected value into account. The alternative is simple to apply and reduces to the Li and Ma method if the systematic error on the normalization factor is negligible. 
\section{Acknowledgements}
The author acknowledges valuable discussions with Jan Conrad. The research of GS is supported by a grant of the Swedish
Research Council (PI: Jan Conrad).
\section*{References}

\bibliography{gerrit_significance}

\end{document}